# A TYPICAL MODEL AUDIT APPROACH
*Spreadsheet Audit Methodologies in the City of London*

Grenville J. Croll
*EuSpRIG - European Spreadsheet Risks Interest Group, grenville@croll-management.freeserve.co.uk*

Abstract:   **Spreadsheet audit and review procedures are an essential part of almost all City of London financial transactions. Structured processes are used to discover errors in large financial spreadsheets underpinning major transactions of all types. Serious errors are routinely found and are fed back to model development teams generally under conditions of extreme time urgency. Corrected models form the essence of the completed transaction and firms undertaking model audit and review expose themselves to significant financial liability in the event of any remaining significant error. It is noteworthy that in the United Kingdom, the management of spreadsheet error is almost unheard of outside of the City of London despite the commercial ubiquity of the spreadsheet.**

Key words:    Spreadsheet Model Risk Audit Review Error Integrity

## 1. INTRODUCTION

The salient characteristics of major financial transactions presently completed in the City of London and elsewhere are invariably modelled in large Microsoft Excel spreadsheets.

The financial magnitude of these transactions in Sterling is typically large - routinely hundreds of millions, often billions and occasionally much larger still. Clearly, with such significant value at stake, it is hardly surprising that the issue of spreadsheet error was recognised and dealt with at an early stage in the City of London.

The two original spreadsheet error detection packages, Cambridge Spreadsheet Analyst and Spreadsheet Auditor were certainly in use in the City of London during the mid eighties. Specialist businesses dedicated to spreadsheet based financial modelling, including the detection and correction of spreadsheet error were established by the early nineties. Specialist financial modelling teams with a model audit and review capability presently exist within the four large professional services firms, and some of the smaller firms, operating in the City.

The terms model audit and model review are used interchangeably. The former tends to be avoided by the larger firms due to liability issues and possible confusion with the corporate audit function. Given the exhaustive nature of the process and the financial liabilities of non-performance, the term model audit would, however, seem to be the most appropriate.

This article describes a spreadsheet model audit process typical of that presently used in the City of London.

## 2. MODEL AUDIT PROCESS

Spreadsheet models typically arrive for audit a few days prior to a transaction's financial completion. The audit process is one of the last procedures during a transaction and tends to get overlooked during the wider activities occurring at an earlier stage. This necessarily places constraints upon the order in which the various phases of model audit can take place.

After a brief familiarisation activity where the structure of the model in terms of sheets, linked sheets and associated databases is appraised, the model is partitioned and dispatched to one or more reviewers for detailed low level scrutiny. Model review often takes place in parallel, where different parts of the model are reviewed simultaneously. Model review is typically performed by graduates with or working for a formal accounting qualification.

Panko [1] has demonstrated that independent cell by cell inspection is almost the only process capable of systematically discovering errors. This is the approach adopted, often with the assistance of proprietary software tools. Model errors are fed back to the developer for correction using a simple error management system.

Microsoft Excel spreadsheets for review are invariably several megabytes with 10Mb not uncommon and 100Mb not unknown. Numbers of unique formulae for inspection are in the range 1,000-10,000 and upwards.

The time taken to review these models can range from twenty five hours to many hundreds, generating significant fee income for firms undertaking this work.

After the low level review has established the correctness of the detailed model formulae and other model parts, a high level review takes place. During the high level review, wider issues such as the correct handling of interest, tax and other financial and accounting issues is determined.

Once the model is correct, it can then be used to investigate sensitivities. That us to say, a few key model variables such as interest rates, cost and revenue assumptions are changed to review the characteristics of the transaction under various commercial scenarios. The performance of the model under these scenarios is checked and documented.

After numerous feedback iterations with the model developer have taken place to correct errors, and with financial closure invariably pressing, a report is issued to the client confirming that various previously agreed review procedures have been performed upon the model. Any unresolved issued are documented.

Financial transactions do not complete until the model audit has been completed. Any delay delays the transaction. Serious model problems have been known to cause transactions to collapse. The legal documentation supporting the transaction reflects the model and not necessarily vice versa. It is increasingly common for the model as audited and agreed at financial close to be the legally agreed tool for monitoring and controlling some key aspects of the deal post financial close. Covenants and restrictions regarding drawdown, repayments and distributions are often enforced via tests run using the original financial model, possibly over a twenty plus year loan life.

## 3. LOW LEVEL REVIEW

A variety of commercial and proprietary software tools exist to determine the likelihood and severity of error in client spreadsheets and the likely location of errors. Most review teams use a variety and combination of tools such as OAK, Space, Spreadsheet Detective and Spreadsheet Professional (reviewed by Nixon and O'Hara in [2]) to increase the chance of error discovery.

Discoverable model characteristics such as formula length, ratio of original to repeated cells, numbers of cell precedents and dependents and the locality and non-locality of cell linkages can be used to infer information about the relative ease or difficulty and time to review a given model.

High level maps of a spreadsheet where the contents of each cell are denoted by a single character such a L for label cell and F for formula cell etc. are commonly used to help ensure that every single cell is examined.

Considerable attention is devoted to ensuring the correctness of ranges and range names, including the following of range naming conventions. Typical problems include overlapping ranges, empty ranges and ranges which do not cover the intended set of cells. Establishing the correctness of ranges early in the low level review process assists with later work.

During the process of examining the formulae in a spreadsheet a large number of issues are considered. These issues can include: ensuring all references to other cells are correct; arguments of functions (such as IF, INDEX, SUM, VLOOKUP etc) are correct; correct use of the ROUND functions; absence of technical errors such as #ERR, #REF etc; absence of circularity; absence of embedded constants; consistency of units of weight and measure; absolute and relative cell addresses and the correct accounting sign of the intermediate and final results.

The documentation trail for a model audit can include a printout of a high level map for the entire spreadsheet with every examined cell ticked and each sheet numbered, signed and indexed.

Most financial spreadsheets contain additional Visual Basic and Macro code which must also be checked. This is done by printing out, inspecting, running and testing all executable entities. This part of the process of reviewing a spreadsheet is directly comparable with the code review phase of traditional software engineering. Grossman [3] deals more fully with the comparisons between software and spreadsheet engineering.

## 4.    HIGH LEVEL REVIEW

Having established the integrity of the low level formulae and code within a model, a high level review can take place. The high level review takes place to check aspects of the overall integrity of the model that may have been missed by the low level review.

A high level review will generally address the consistency of the model with any supporting documentation. This part of the review will also include checks to ensure that finance and accounting issues have been dealt with correctly. That is to say that the model follows Generally Accepted Accounting Practices (GAAP) relevant to the jurisdiction within which the transaction takes place.

Depending upon which firm is performing the model audit, the high level review may or may not address commercial issues relevant to the transaction. Often it is too late to address issues such as the viability or commercial logic of the transaction and its terms. It is often the case, however, that the true nature of the transaction only becomes visible once a correct model is in place.

Typical high level checks for a financial model cover issues such as: ensuring the balance sheet balances; checking whether retained earnings flow from the profit and loss account to the balance sheet; ensuring debt is amortised correctly; ensuring fixed assets do not depreciate below zero;

checking whether revenues and costs reflect production; ensuring tax and deferred tax is handled correctly and so forth.

High level checks on the model and its documentation will certainly include a detailed review of the term sheet of any debt or other funding to ensure that the precise details of its provision are reflected in the model. The funding provisions can be very detailed, particularly where financial reserves are accumulated to ensure debt is serviced correctly. Other documentation checks can include detailed matching of model constants such as revenue and cost estimates to the original documents containing them and vice versa to ensure that the model contains the required detail.

Any changes to the model at the high level review stage cause a re-review of the model at low-level. Comparison software is often used to assist in locating model changes between versions.

## 5.     SENSITIVITIES

By now, the model is largely correct. That is to say, if a specification existed (more often than not, they do not), the model would largely perform to that specification.

Running sensitivities involves running the model with a handful of key variables changed either singly or in combination. Key variables include revenue & costs - both high and low, start date, interest rates, discount rates and debt/equity ratios. For a large model, running sensitivities can be time consuming with model re-calculate times of several hours not being unknown. For a well designed and constructed model, the sensitivities will run without error, yielding financial information essential to the completion of the transaction. Errors do occur at this late stage, requiring rapid repair and re-review prior to re-execution of all the sensitivities.

The results of each sensitivity are methodically reviewed and documented. A version of the model containing the changed variables will be saved.

## 6.     FINAL REVIEW & REPORT

The concluding part of a model audit is a drawing together of all the work that has been performed upon the model. A package of documentation will be assembled which can be used (in court if necessary) to prove that every check and test that was required to be performed was performed.

Any final queries regarding model performance will be cleared if possible. Any areas of concern that remain will be clearly documented.

The wording of the final report to the client will vary according to which firm is performing the work, what type of work was performed and who the client is. Wording can vary from simply stating that various agreed upon procedures have been performed through to statements that the model is free of material error.

Care is taken in final reports to identify exactly which model was the subject of the audit. Typical identification data will include basic information such as file name, date, time and size.

The limit to the contractual liability of the firm performing the model audit work is an important issue. A liability figure is always agreed prior to any work taking place and often re-stated in the final report. In some cases liability is, or has to be, unlimited, placing onerous responsibilities upon firms and individuals that perform this work on large transactions.

## 7. CONCLUSION

Model audit is a detailed, time consuming and essential prerequisite for the consummation of financial transactions in the City of London. Correctness of spreadsheet models ensures that the parties to a transaction can rely upon the integrity of the financial information presented.

The methodologies used for ensuring the correctness of spreadsheet models are generally straightforward and have parallels in more widely studied aspects of software engineering, where code review is the norm.

Analysis of the client bases of firms performing model audit suggests that model audit is largely confined to financial work performed in the main financial centres. Given the ubiquity of the spreadsheet and spreadsheet error, the integrity and reliability of spreadsheets outside of this narrow field is questionable.

## 8. ACKNOWLEDGEMENTS

The author thanks David Chadwick, Ray Butler and Penny Lynch for their constructive comments upon a draft of this paper.